\documentclass[epj]{svjour}
\usepackage{graphics,amssymb}
\usepackage{color}
\usepackage{xcolor} 
\newcommand{\grb}[1]{\mbox{\boldmath $#1$}}
\begin{document}
\title{Transport and selective chaining of bidisperse particles
in a travelling wave potential}
\titlerunning{Transport and selective chaining of bidisperse particles in a travelling wave potential}
\author{Pietro Tierno\inst{1,2} \and Arthur V. Straube\inst{1}}
\authorrunning{P. Tierno and A. V. Straube}
\institute{Departament de F\'isica de la Mat\`eria Condensada,
Universitat de Barcelona, Av. Diagonal 647, 08028 Barcelona, Spain \and
Institut de Nanoci\`encia i Nanotecnologia IN$^2$UB,
Universitat  de Barcelona, Barcelona, Spain}
%
\date{}
\abstract{We combine experiments, theory and
numerical simulation to investigate the dynamics of a binary
suspension of paramagnetic colloidal particles dispersed
in water and transported above a stripe patterned magnetic garnet film.
The substrate generates a one-dimensional periodic
energy landscape above its surface. The application of an elliptically polarized
rotating magnetic field causes the landscape to translate, inducing direct
transport of paramagnetic particles
placed above the film.
The ellipticity of the
applied field can be used to control
and tune the interparticle
interactions,
from net repulsive to net attractive.
When considering particles of two distinct sizes,
we find that, depending on their elevation above the
surface of the magnetic substrate, the particles feel
effectively different potentials, resulting in different
mobilities.
We exploit this feature to induce
selective chaining for certain values of the applied field parameters.
In particular, when driving two types of particles,
we force only one type to condense into travelling
parallel chains. These chains confine the
movement of the other non-chaining
particles within narrow colloidal channels.
This phenomenon is explained by
considering the balance of
pairwise magnetic forces between the particles
and their individual coupling with the travelling landscape.  }
%
\maketitle
\section{Introduction}

Recent years have witnessed
an increasing interest in
developing novel
techniques
which make use of uniform
magnetic field
modulated by a periodic substrate
in order
to induce the controlled motion of
colloidal microspheres
in water~\cite{Mir05,Ram06,Gun05,Yel05,Tie07,Don10,Hen11,Ehr11}.
In contrast to
optical or electric field micromanipulation,
magnetic fields have the advantages that
they neither alter the fluid medium nor affect
biological systems, although their use is limited to
polarizable particles~\cite{Gijs10,Mar13}.
Magnetophoresis, \textit{i.e.} the controlled
transport of particles
via an external field gradient, is a well established
method with several applications in
biomedical research and clinical diagnostics~\cite{Zbo99,Suw11}.
However, precise control of particle position and speed in a single chip
is difficult to obtain with
a field gradient since the amplitude varies spatially over
an extended area.

Magnetic fields that are heterogeneous on the particle scale
can guarantee a precise and selective
manipulation of both individual and large collection
of colloidal microspheres.
Such fields can be obtained
by using magnetic patterned substrates
which contain, for example,
permalloy islands~\cite{Gun05},
cobalt microcylinders~\cite{Yel05}, domain
wall conduits~\cite{Don10}, magnetic wires~\cite{Hen11}, or
even exchange bias systems~\cite{Ehr11}.
Another method consists in using
ferrite garnet films (FGFs), \textit{i.e.} epitaxially grown
single crystalline films
where magnetic domains
organize into
patterns of stripes
with a spatial periodicity of few microns.
These domains generate a
one dimensional periodic
potential which can be used to trap~\cite{Hel05},
assemble~\cite{Tie08} or transport~\cite{Tie10}
paramagnetic colloidal particles
deposited above the film.
In the latter case,
it was found that an external
rotating magnetic field is able to create
a moving landscape, similar to a travelling wave potential,
which can transport the particles at a well defined speed~\cite{Tie20122}.
In particular, depending on the driving frequency, two dynamic states
are possible: (i) at low frequencies, the particles
are synchronized with the external field and are therefore
transported with the speed of the travelling landscape;
(ii) beyond a critical
value $f_{\rm c}$, the particles desynchronize with the translating potential,
showing a complex sliding dynamics characterized
by a global decrease of their average speed.
However, theoretical arguments~\cite{Str2013} reveal
that the transition between both states is strongly sensitive to
the particle elevation above the substrate,
opening the possibility to separate
magnetic particles based on their relative size.
This feature was already demonstrated in different
works using similar~\cite{Tie081,Tie082},
or different~\cite{Yel07} magnetic substrates.
However, the role of interparticle interaction in this process
and their relative effect on the collective dynamics for a
bidisperse colloidal system has not been explored so far.

In this article, we study the
dynamics and interactions between paramagnetic colloidal particles
transported above a travelling wave potential.
We focus on a binary mixture
where different elevation above the magnetic substrate
modifies the particle collective behaviour under an applied field.
In particular, an elliptically polarized
magnetic field
is used to induce
strong attraction between
one type of particle,
triggering the formation of long
chains travelling at constant speed along the corrugated potential.
\begin{figure*}
\begin{center}
\resizebox{\textwidth}{!}{\includegraphics{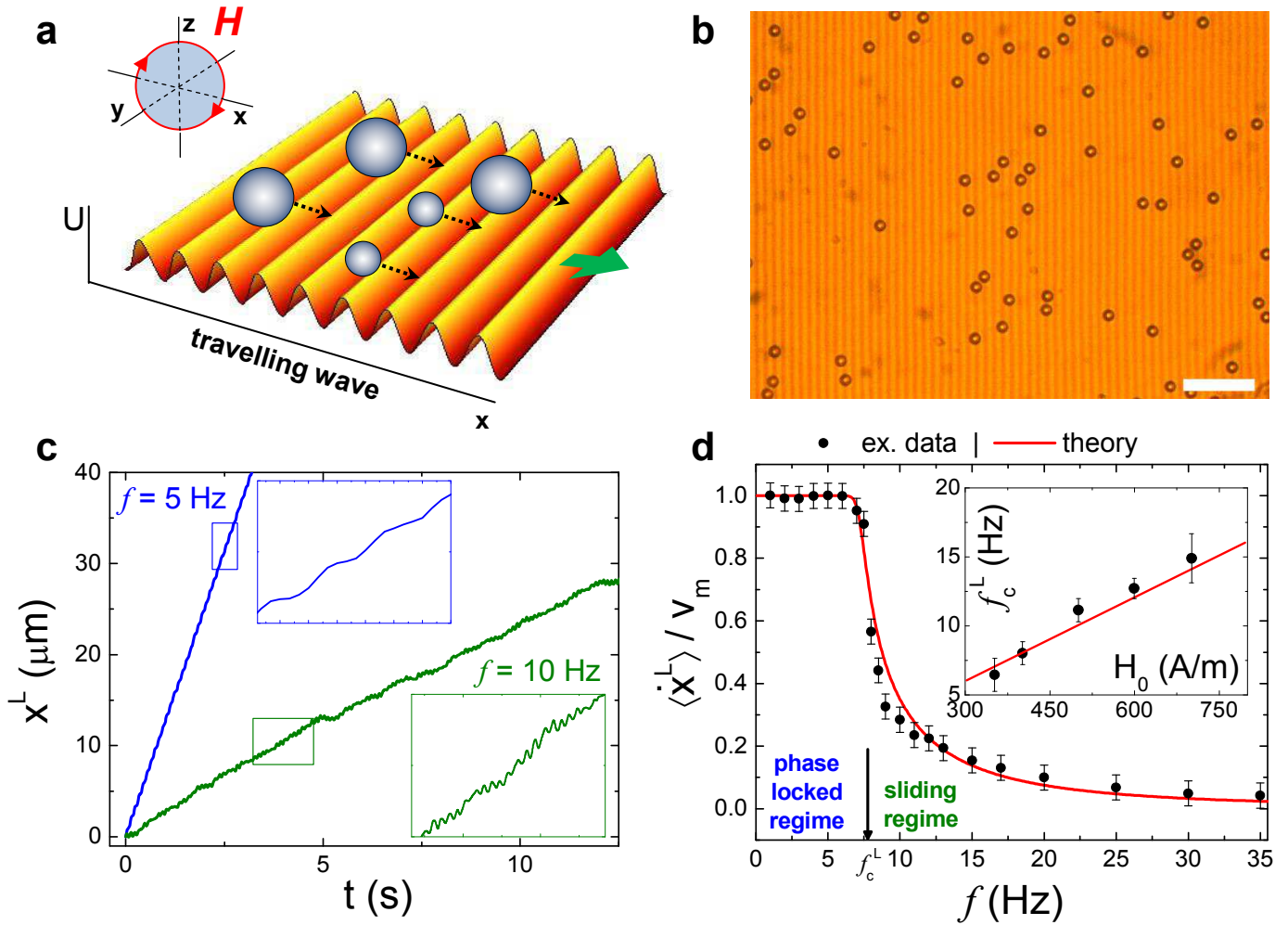}}
\caption{(a) Schematic showing a binary mixture of paramagnetic colloidal particles
transported above a travelling periodic energy landscape. The latter is induced
by an external magnetic field
rotating in the $(x,z)$ plane.
(b) Polarization microscope image
showing the stripe patterned
FGF (of spatial periodicity $\lambda = 2.6 \,{\rm \mu m}$)
with a diluted sample of large (${\rm L}$)
particles (diameter $d^{\rm L}=2.8 \, {\rm \mu m}$).
Scale bar is $20 \, {\rm \mu m}$.
(c) Trajectory of one large particle
driven across the FGF by an applied circularly polarized ($\beta=0$) field
with amplitude $H_0=400\, {\rm A/m}$
and at two frequencies $f=5 \,{\rm Hz}$ (blue, phase locked regime)
and  $f=10 \,{\rm Hz}$ (green, sliding regime).
(d) Particle velocity $\langle \dot{x}^{\rm L} \rangle /v_{\rm m}$ normalized
by the speed $v_{\rm m}=\lambda f$ of translation of the landscape
as a function of driving frequency $f$ for a large paramagnetic
colloidal particle subjected to a rotating field with amplitude
$H_0=400 \, {\rm A/m}$ and ellipticity $\beta=0$.
Continuous line denotes the nonlinear fit following
Eq.~(\ref{speed-stoch-beta=0}) in the text, while scattered points (filled circles) are
experimental data. Inset shows the
dependence of the critical frequency $f^{\rm L}_{\rm c}$
as a function of the field amplitude $H_0$.}
\label{fig:1}
\end{center}
\end{figure*}
However, this chaining effect does not occur for the
other colloidal species,
which slides close to the moving chains.
We explain this peculiar behaviour by using
simple theoretical arguments, and
confirm the observed dynamics
via numerical simulations.

\section{Experimental system}
\label{Exp}
The substrate potential is generated by
using an FGF
of thickness $\simeq 4 \,{\rm \mu m}$
grown by dipping liquid phase epithaxy~\cite{TiePCCP}.
The FGF is characterized by a stripe pattern of ferromagnetic domains
having alternating magnetization
with a spatial periodicity
given by $\lambda=2.6 \pm 0.2 \, {\rm \mu m}$, Fig.~\ref{fig:1}(b).
These domains are
separated by Bloch walls (BWs),
\textit{i.e.} transition regions where the
magnetization vector rotates by $180$ degrees
out of the film plane,
and the stray field of the film is maximal.
Because of this arrangement of the magnetic domains,
under no external field
the FGF generates on its surface a
static spatially periodic potential,
which is characterized by a step-like profile directly
at the surface of the FGF and a smoothed sinusoidal-like profile far away from the surface.
We transport above the FGF two types
of paramagnetic colloidal particles,
having diameters
$d^{\rm S}=1.0 \, {\rm \mu m}$ (Dynabeads myone, Dynal)
and $d^{\rm L}=2.8 \, {\rm \mu m}$ (Dynabeads M-270, Dynal),
and effective magnetic volume susceptibility
$\chi^{\rm S} \simeq 1$~\cite{Cli2007} and
$\chi^{\rm L} \simeq 0.4$~\cite{Hel2007},
respectively. Hereafter, the superscripts ``{\rm S}'' and ``{\rm L}'' are used to refer to
the small and large particles.
Both types of particles
are composed of a cross-linked polystyrene
matrix doped with superparamagnetic iron oxide grains
and coated with COOH surface groups.
The particles are diluted
with deionized water (milli-Q, Millipore)
and deposited above the FGF,
where they sediment due to density mismatch.
Once close to the surface of the FGF,
the particles are attracted to
the BWs which
confines their motion in two dimensions.
In order to decrease the strong attraction of
the BWs and avoid particle sticking,
the FGF is coated with a $1\, {\rm \mu m}$
thick layer of a photoresist (AZ-1512 Microchem, Newton,
MA) using spin coating and UV-photo crosslinking~\cite{Tie2012}.
Thus the effective particle
elevation above the surface of the
FGF (\textit{i.e.} the shortest distance from the surface of the
FGF to the center of particle)
is $z^{\rm S}=1.5 \, {\rm \mu m}$ and $z^{\rm L}=2.4 \, {\rm \mu m}$
for the small and large particle, respectively.

The particle dynamics are observed
by using a polarization
optical microscope (Eclipse Ni, Nikon)
which is equipped with a $100 \times \, 1.3$ NA
microscope objective
and a $0.45$ TV lens.
The microscope has a CCD camera
(Basler Scout scA640-74)
which is capable of recording real-time video clips of the particles
up to $75$ frames/s in a total field of view
of $140 \times 105 \, {\rm \mu m^2}$.
We extract the particle trajectories
from these recorded data using a
custom made tracking software
based on the Crock and Grier
original code~\cite{Cro96}.
The external field is applied by using
a custom made magnetic coil system
mounted on the stage of the optical microscope
and connected to
two independent power amplifiers (BOP 20 10-M, KEPCO)
driven by an
arbitrary waveform generator (TGA1244, TTi).

\section{Theoretical model}

Consider a mixture of spherical paramagnetic particles numbered by $l=1,2,3, \dots$.
Because the particles are bidisperse, they are characterized by
diameters $d_l=\left\{d^{\rm S},d^{\rm L}\right\}$ and the effective
volume susceptibilities $\chi_l=\left\{\chi^{\rm S},\chi^{\rm L}\right\}$,
as described in Sec.~\ref{Exp}. A paramagnetic spherical particle $l$ positioned at ${\mathbf r}_l$
and subjected to an external field ${\mathbf H}$ acquires a dipole
moment ${\mathbf m}_l=V_l \chi_l {\mathbf H}_l$ \cite{Car15},
where $V_l=\pi d_l^3/6$ is the volume of particle and the field
is taken at the position of particle, ${\mathbf H}_l = {\mathbf H}({\mathbf r}_l,t)$.
This dipole interacts with the external magnetic field and with other
induced dipoles, say dipole $l'$ at a position ${\mathbf r}_{l'}$, whose interaction potentials can be written as
\begin{eqnarray}
U_{\rm s}(\mathbf{r}_l,t) & = & -\frac{1}{2} V_i \chi_l \mu_{\rm 0} {\mathbf H}^2(\mathbf{r}_l,t)  = -\frac{1}{2} V_i \chi_l \mu_{\rm 0} {\mathbf H}^2_l,  \label{Us-gen}\\
U_{\rm dd}(\mathbf{r}_{ll'},t) & = & \frac{\gamma_{ij}}{r_{ll'}^3} \left[ ({\mathbf H}_l \cdot {\mathbf H}_{l'})-3 ({\mathbf H}_l \cdot \hat{\mathbf r}_{ll'}) ({\mathbf H}_{l'} \cdot \hat{\mathbf r}_{ll'}) \right], \label{Udd-gen}
\end{eqnarray}
respectively. Here, $\mu_{\rm 0}=4\pi\times 10^{-7}\,{\rm H \, m^{-1}}$ is the magnetic
permeability of the free space, $\gamma_{ll'}= \chi_l \chi_{l'} V_l V_{l'} \mu_0/(8\pi)$,
${\mathbf r}_{ll'}={\mathbf r}_l-{\mathbf r}_{l'}$, $r_{ll'}=|{\mathbf r}_{ll'}|$, and
$\hat{\mathbf r}_{ll'}={\mathbf r}_{ll'}/r_{ll'}$.

For our system, the magnetic field $\mathbf{H}$ is the total field
above the FGF, given by the superposition $\mathbf{H}=\mathbf{H}^{\rm sub}+\mathbf{H}^{\rm ac}$
of the stray field of the substrate, $\mathbf{H}^{\rm sub}$, and the field of modulation, $\mathbf{H}^{\rm ac}$.
To a good accuracy, the substrate field $\mathbf{H}^{\rm sub}$ can be approximated as~\cite{Str2013}
\begin{equation}
\mathbf{H}^{\rm sub}(\mathbf{r})=\frac{4 M_{\rm s}}{\pi} {\rm e}^{-kz}\left(\cos kx,0,-\sin kx \right),
\label{Hsub}
\end{equation}
where $M_{\rm s}$ denotes its saturation magnetization and $k=2\pi/\lambda$ is the wavenumber. The superimposed ac field rotates in the $(x,z)$ plane and has elliptic polarization:
\begin{equation}
{\mathbf H}^{\rm ac}(t)=(H_{0x} \cos \omega t, 0, -H_{0z} \sin \omega t),
\label{Hac}
\end{equation}
where $\omega=2\pi f$ is the angular frequency. The amplitude of modulation $H_0$ and the
ellipticity parameter $\beta \in [-1,1]$ are introduced as~\cite{Lacis1997}
$H_0=\sqrt{(H_{0x}^2+H_{0z}^2)/2}$ and $\beta=(H_{0x}^2-H_{0z}^2)/(H_{0x}^2+H_{0z}^2)$.
The partial case of $\beta=0$ corresponds to the case of circular polarization, $H_{0x}=H_{0z}$.

The dynamics of particles is considered overdamped and two dimensional, at a fixed
elevation above the FGF, resulting in the following Langevin equations
\begin{equation}
\zeta_l \frac{d {\mathbf r}_l}{dt} = {\mathbf F}_{\rm s}({\mathbf r}_l,t) + \sum_{l'} {\mathbf F}_{\rm d}({\mathbf r}_{ll'},t) + \sqrt{2 \zeta_l k_{\rm B} T } \,\grb{\xi}_l(t)\,, \label{LEs} 
\end{equation}
where $\zeta_l=3\pi \eta d_l$ is the viscous friction
coefficient ($\eta$ is the dynamic viscosity of the solvent),
${\mathbf F}_{\rm s}({\mathbf r}_l,t)=F_{\rm s}(x_l,t)\hat{\mathbf e}_x$
($\hat{\mathbf e}_x$ is the unit vector of the $x$ axis), $F_{\rm s}(x_l,t)=-\partial_{x_l}U_{\rm s}({\mathbf r}_l,t)$
is the force exerted on particle $l$ by the external field, and ${\mathbf F}_{\rm dd}({\mathbf r}_{ll'},t)=-\partial_{{\mathbf r}_l}U_{\rm dd}({\mathbf r}_{ll'},t)$
is the dipolar force from particle $l'$. The last term in Eq.~(\ref{LEs}), in which $k_{\rm B}T$ is the thermal energy,
is the stochastic force taking account of thermal fluctuations modelled by the Gaussian white noise with the mean
and covariance given by $\left<\grb{\xi}_{l'}(t)\right>=0$
and $\left<\grb{\xi}_l(t)\grb{\xi}_{l'}(t')\right>=\mathbf{I}\,\delta_{ll'}\delta(t-t')$,
respectively. Here, $\mathbf{I}$ is the second-order identity tensor.

To account for the finite size of particles, we also include hard-core repulsive potential,
as we did in Ref.~\cite{Str2014}.

\section{Discussion}

\subsection{Individual particle propulsion for circular polarization}

In the case of circular polarization, $\beta=0$, the potential
that describes the single particle motion, $U_{\rm s}$ as in
Eq.~(\ref{Us-gen}), evaluated based on expressions (\ref{Hsub}) and
(\ref{Hac}) corresponds to a sinusoidal wave propagating with the speed $v_{\rm m}(f)=\lambda f$ across
the stripes of the FGF, $U_{\rm s}(x_l,t)\propto \cos[k(x_l-v_{\rm m} t)]$.
As a result, $v_{\rm m}$ is the maximum speed that the particles acquire
when following the potential at low frequencies.
This phase-locked motion is characterized by a constant
propulsion speed, as confirmed by an almost linear
particle trajectory shown by the blue line in Fig.~\ref{fig:1}(c).
At high frequencies, particles decouple
from the potential, moving in an asynchronous way with on
average smaller speeds. Now the particles
are in a sliding regime,
displaying a series of oscillations along their
motion, see the green curve in Fig.~\ref{fig:1}(c). Both regimes of motion can be
characterized in terms of the average propulsion speed, $\left<\dot x_l\right>$,
as shown in Fig.~\ref{fig:1}(c) for a single particle of size $d^L$.

It can be shown that the deterministic ($T=0$)
averaged speed of the particles can be derived as,~\cite{Str2013}
\begin{equation}
\left<\dot x_l\right>_{\beta=0} = v_{\rm m} \left\{
\begin{array}{ll}
1, & {\rm if} \;\; {f} < {f}_{{\rm c}l}(0)\,, \\
1 - \sqrt{1-{f}_{{\rm c}l}^2(0)/{f}^2 }\,, & {\rm if} \;\; {f} > {f}_{{\rm c}l}(0)\,, \\
\end{array}
 \right.
\label{speed-det-beta=0}
\end{equation}
where
\begin{equation}
f_{{\rm c}l}(\beta=0)=\frac{8M_{\rm s} H_0 \mu_0 V_l \chi_l}{\zeta_l \lambda^2}{\rm e}^{-kz_l} \label{f-crit-beta=0}
\end{equation}
is the critical frequency, as denoted by the subscript ``c''.
The subscript ``$l$'' indicates that the critical frequency is generally
particle dependent since particles of different sizes are characterized by
different values of the parameters $V_l$, $\chi_l$, $\zeta_l$, and $z_l$.

In contrast to a sharp transition displayed by the deterministic prediction, Eq.~(\ref{speed-det-beta=0}),
the account of thermal fluctuations ($T>0$) smoothens the crossover from the phase-locked to
the sliding motion close to the critical point. Thus the effect of thermal noise on the average
particle speed is described by
\begin{equation}
\left<\dot x_l\right>_{\beta=0} = v_{\rm m} \left[1-\frac{\sinh(\pi \alpha_l)}{\pi \alpha_l \,|I_{{\rm i} \alpha_l}(\alpha_{{\rm c}l})|^2}\right],
\label{speed-stoch-beta=0}
\end{equation}
%
where $I_{{\rm i}\nu}(x)$ is the modified Bessel function
of the first kind of an imaginary order, and we have introduced dimensionless parameters
\begin{equation}
\alpha_l=\frac{\zeta_l f \lambda^2}{2\pi k_{\rm B}T}\,, \quad  \alpha_{{\rm c}l}=\frac{\zeta_l f_{{\rm c}l}(0) \lambda^2}{2\pi k_{\rm B}T}\,. \label{alphas}
\end{equation}
Equation (\ref{speed-stoch-beta=0}) is used to fit the experimental data for the average speed $\left<\dot x_l\right>$
for the large particles with $d_l=d^{\rm L}$ at $H_0 = 400 \,{\rm A/m}$
against the theoretical predictions, see Fig.~\ref{fig:1}(d). We find
that at this amplitude of the applied field, the critical frequency
is given by $f_{\rm c}^{\rm L}(0)=7.7 \, {\rm Hz}$,
which allow us to estimate the saturation magnetization
$M_{\rm s} \approx 24600 \, {\rm A/m}$, used as a fitting parameter.
Note that the linear dependence of the critical frequency
on the amplitude of modulation can be seen from the inset of Fig.~\ref{fig:1}(d),
which is in accordance with the theoretical prediction given by Eq.~(\ref{f-crit-beta=0}).

\begin{figure}
\begin{center}
\resizebox{\columnwidth}{!}{\includegraphics{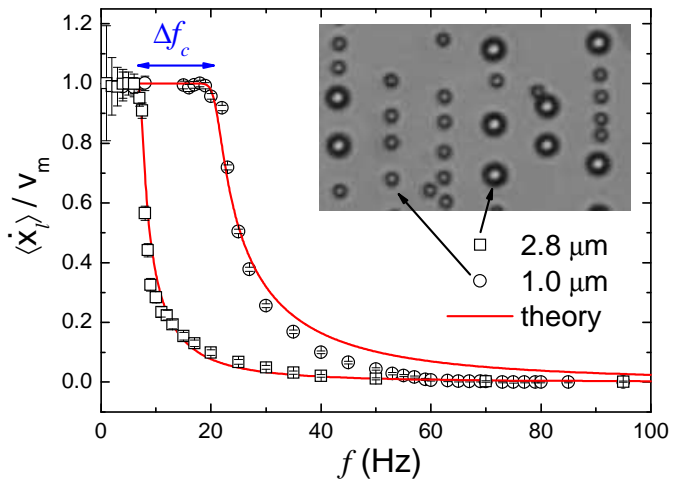}}
\caption{Normalized particle velocity $\langle \dot{x} \rangle /v_{\rm m}$ (where $v_m=\lambda f$) as
a function of the driving frequency $f$ for a large ($2.8 \,{\rm \mu m}$,
empty squares)
and small ($1.0 \,{\rm \mu m}$, empty circles)  paramagnetic
colloidal particle subjected to a rotating field with amplitude
$H_0=400 \,{\rm A/m}$ and ellipticity $\beta=0$. The continuous red lines
are fits according to Eq.~(\ref{speed-stoch-beta=0}),
and $\Delta f_{\rm c} = f^{\rm S}_{\rm c}-f^{\rm L}_{\rm c}=12.6 \, {\rm Hz}$ denotes the difference between the critical
frequencies of the large (${\rm L}$) and the small (${\rm S}$) particle.
Inset shows a microscope image
of both kinds of particles above the FGF.
The vertical stripes are not visible due to the absence of polarization elements.}
\label{fig:2}
\end{center}
\end{figure}

We next show in Fig.~\ref{fig:2}
the normalized mean speeds as a function of frequency for
particles with the diameters $d^{\rm L}$ and $d^{\rm S}$
and subjected to a rotating magnetic field
with amplitude $H_0 = 400 \,{\rm A/m}$.
As expected, given the different values of the particle parameters,
the smaller particles ($d^{\rm S}$) require
a much higher frequency barrier to desynchronize
with the travelling potential and enter the sliding regime.
The corresponding curve (empty circles in Fig.~\ref{fig:2})
can be well fitted by Eq.~(\ref{speed-stoch-beta=0})
using the same value of $M_{\rm s}$, as in Fig.~\ref{fig:1}(d).
We find the critical frequency for the small particles
to be $f^{\rm S}_{\rm c}=20.3 \,{\rm Hz}$, which is so high
that the average speed of the large particles
has already decreased by $90\%$,
$\langle \dot x^{\rm L}\rangle=0.1 \, v_{\rm m}$.
This feature enables us
to separate the paramagnetic particles based
on the significant difference in their relative speeds,
when driving both types of particles
at a frequency above $f^{\rm L}_{\rm c}=7.7 \,{\rm Hz}$.
For frequencies higher
than $100 \,{\rm Hz}$ the particles eventually
slow down till to be
practically unable to follow the quickly
moving landscape.
In contrast, below $f^{\rm L}_{\rm c}=7.7 \,{\rm Hz}$, both types of particles
can be transported at the same constant speed, $v_{\rm m}$,
staying completely localized along
a series of equipotential lines of the travelling potential, equally spaced and aligned along the stripes,
as shown in the small inset in Fig.~\ref{fig:2}.

\subsection{Interacting particles for elliptic polarization. Reduced model}

When transported by the travelling potential,
the paramagnetic colloidal particles also interact with each other
because of magnetic dipolar forces.
It has recently been shown in a previous work~\cite{Str2014}
that these forces can be tuned by varying the ellipticity $\beta$ of the rotating field.
A pair of similar particles moving one behind another
above the FGF display either repulsive or attractive
interactions for an ellipticity parameter $\beta<\beta_{\rm c}$ or $\beta>\beta_{\rm c}$,
respectively. Here $\beta_{\rm c}=-1/3$ denotes the transition
between both types of behaviour.
For ellipticity $\beta$ significantly larger than $\beta_{\rm c}$,
strong dipolar interactions force the particles to rapidly self-assemble into
travelling chains.
The interpretation of chaining for the general case of
elliptically polarized field, $\beta \ne 0$, becomes straightforward
in terms of reduced equations of motion.

In this case, the propulsion of particles can effectively be described by a time-independent single-particle potential
%
$U_{\rm s}^{\rm eff}(x_l)=\zeta_l \left<\dot x_l\right> x_l$,
%
which is linearly dependent on the coordinate $x_l$ and describes the individual propulsion
with constant speed ${\mathbf v}_{\rm s}({\mathbf r}_l) = \left<\dot x_l\right> \hat{\mathbf e}_x$
along the $x$ axis. The speed of propulsion depends on the frequency $f$, the ellipticity of the
field, $\beta$, and the size of particle. Similarly to the case $\beta=0$, at a low frequency
$f<f_{{\rm c}l}(\beta)$, the particle propels with the maximum speed, $\left<\dot x_l\right>=v_{\rm m}$.
At high frequencies, the speed of propulsion drops down. Generally, the value of $f_{{\rm c}l}(\beta)$
and the dependence of $\left<\dot x_l\right>$ on $f$ and $\beta$ can be obtained only numerically.
However, for frequencies $f$ that are formally far beyond the critical value $f_{{\rm c}l}(\beta)$,
which represents a good approximation
already for $f \gtrsim 2 f_{{\rm c}l}(\beta)$, we can apply the estimate
$\left<\dot x_l\right>_{\beta\ne 0}=\left<\dot x_l\right>_{\beta = 0} \sqrt{1-\beta^2}$~\cite{Str2014}.
Here, $\left<\dot x_l\right>_{\beta=0}=(v_{\rm m}/2)f_{{\rm c}l}^2(0)/f^2$,
which coincides with Eq.~(\ref{speed-det-beta=0}) considered for
$f \gg f_{{\rm c}l}(0)$. These asymptotic results allow us to outline a rough estimate for
the critical frequency at nonzero ellipticity, which would be given by
$f_{{\rm c}l}(\beta)\approx f_{{\rm c}l}(0) (1-\beta^2)^{1/4}$.

For the dipolar interaction of particles, the full potential as in Eq.~(\ref{Udd-gen}) can
be replaced by its reduced counterpart, see Eq.~(34) in Ref.~\cite{Str2014}:
\begin{equation}
U_{\rm dd}^{\rm eff}({\mathbf r}_{ll'})=\frac{\gamma_{ll'} H_0^2}{r_{ll'}^3}\left[1-\frac{3(1+\beta)}{2}\frac{({\mathbf r}_{ll'}\cdot \hat{\mathbf e}_x)^2}{r_{ll'}^2}\right]. \label{Udd-eff}
\end{equation}
Note that this approximation taken at $\beta=0$ is in agreement with
the reduced potential obtained in the context of front propagation for
circularly polarized modulation, see Eq.~(19) in Ref.~\cite{Fer16} at $H_x=0$, $H_y=0$.
This potential is always repulsive for particles moving along the landscape side by side, $x_{ll'}=0$,
and is conditionally attractive for particles moving one behind another, $y_{ll'}=0$, provided that $\beta > -1/3$.

\subsection{Tuning structure formation leading to selective chaining}

We will now consider a situation in which the driving parameters are chosen
in such a way that the small particles are in the subcritical
(phase-locked) regime and the large particles in the supercritical (sliding) regime.
First, let us fix the field amplitude to be $H_0 = 780 \, {\rm A/m}$ and the
ellipticity $\beta=-0.12$. Because the ellipticity is relatively small, $|\beta| \ll 1$,
from our estimates we find that $f_{\rm c}(\beta) \approx f_{\rm c}(0)$. Taking into
consideration Eq.~(\ref{f-crit-beta=0}), at the given value of $H_0$ and the estimate of $M_{\rm s}$
we find the critical frequencies for the large and small particles,
$f_{\rm c}^{\rm L} \approx 15 \,{\rm Hz}$ and $f_{\rm c}^{\rm S} \approx 42 \,{\rm Hz}$,
respectively. Now, by choosing the frequency of modulation to be $f=30 \, {\rm Hz}$, we
ensure that the small particles will propel with the maximum speed, $\left< \dot x^{\rm S} \right>=v_{\rm m}$
because $f < f_{\rm c}^{\rm S}(\beta)$, whereas the large particles are in the sliding regime and will move
much slower on average, since $f > f_{\rm c}^{\rm L}(\beta)$. From Eq.~(\ref{speed-det-beta=0}) and
a high-frequency approximation, we obtain close estimates,
$\left< \dot x^{\rm L} \right> \approx v_{\rm m}-v_{\rm m}\sqrt{1-({f}_{\rm c}^{\rm L}/f)^2}\approx 0.134 \,v_{\rm m}$
and $\left< \dot x^{\rm L} \right> \approx v_{\rm m}({f}_{\rm c}^{\rm L}/f)^2/2 \approx 0.125 \, v_{\rm m}$.
Note that both these analytic estimates are also close to the accurate value, $\left< \dot x^{\rm L} \right> =\approx 0.136 \, v_{\rm m}$, obtained by numerically averaging the speed of a single particle.

We are now ready to show why chaining of only large particles is achieved. As described above,
there are two independent factors that govern the dynamics of particles: the interaction of particles
with the field above the FGF and the dipolar interaction between the particles. The relative contribution
of these factors is drastically different for small and large particles. To validate this statement, we
compare the strength of dipolar interactions relative to the energy of individual interaction with the field
of substrate. As follows from Eq.~(\ref{Udd-eff}), the characteristic energy of dipolar
interaction of a pair of particles moving one behind another,
$y_{ll'}=0$, $r_{ll'}=x_{ll'}$, is $U_{\rm dd}^{\rm eff}(r_{ll'})=-\gamma_{ll'} H_0^2(1+3\beta)/(2r_{ll'}^3)$.
Since the propulsion force is $\zeta_l\left<\dot x_l\right>$, the characteristic energy caused by the
interaction of particle $l$ with the field of substrate can be estimated as $U_{{\rm s}}=\lambda \zeta_l\left<\dot x_l\right>$.
Therefore, for the relative energy of dipolar interaction we obtain
\begin{equation}
\frac{U_{\rm dd}(r)}{U_{\rm s}}=-\frac{\gamma H_0^2(1+3\beta)}{2\lambda\zeta \left<\dot x \right> r^3}\,, \label{Udd-rel}
\end{equation}
where $\zeta$ and $\left<\dot x\right>$ are to be
replaced by either $\zeta^{\rm L}$ and $\left<\dot x^{\rm L}\right>$ for
large or $\zeta^{\rm S}$ and $\left<\dot x^{\rm S}\right>$ for small particles, respectively.

\begin{figure}
\begin{center}
\resizebox{\columnwidth}{!}{\includegraphics{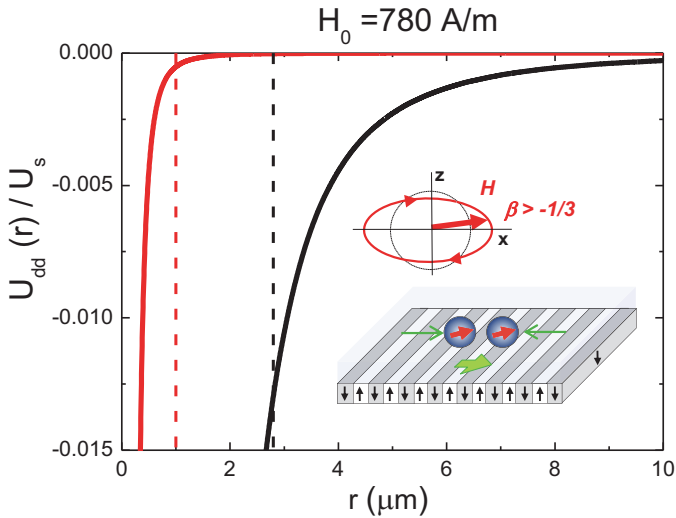}}
\caption{Normalized magnetic dipolar interaction $U_{\rm dd}/U_{\rm s}(r)$
as given by Eq.~(\ref{Udd-rel})
between two small (red line) and two large (black line)
paramagnetic particles when subjected
to an external rotating magnetic field with
amplitude $H_0=780 \, {\rm A/m}$
and ellipticity $\beta = -0.12$.
$U_{\rm s}$ denotes the characteristic energy of interaction of the
particle with the field above the FGF, which is dependent
on whether the particles are in the phase-locked or in the sliding
regime,
see text for details.
Dashed lines denote the corresponding
separation distance between the particles at contact,
equal to $d^{\rm S}$ for small and $d^{\rm L}$ for large particles.
The schematic in the inset illustrates
the geometry considered:
particles having no relative
displacement along the stripes
display attractive dipolar
interactions when
the ellipticity of the rotating field
$\beta > -1/3$.}
\label{fig:3}
\end{center}
\end{figure}

The extreme cases of interactions between large
and large particles (or small and small particles) with $\gamma$
replaced by $\gamma^{\rm LL}$ (or $\gamma^{\rm SS}$) are shown in
Fig.~\ref{fig:3}. The fact that the corresponding energies and hence
the corresponding forces at contact are strongly separated indicates
that only the dipolar interactions between the large particles is non-negligible.
Note that a similar analysis of interactions between small and large particles
shows that although these interactions are stronger than those between the small
particles they are significantly smaller than the dipolar interactions between
the large particles. This fact can be explained by the fact that the interaction
with the field of substrate dominates the force balance for small particles,
which are strongly coupled to the translating substrate potential and the dipolar
coupling to other particles is negligibly small. The strength of interaction of
large particles with the field of substrate is significantly weaker. Therefore
the weak dipolar forces become comparable with the forces responsible for
propulsion and can no longer be neglected.

As a result, the dipolar coupling strengths $\gamma^{\rm SS}$ and $\gamma^{\rm SL}$
can be set to zero and the dynamics of small and large particles is governed by the
effective equations
\begin{eqnarray}
\frac{d {\mathbf r}^{\rm S}_l}{dt} & = & v_{\rm m} \hat{\mathbf e}_x +\sqrt{2 D^{\rm S} } \,\grb{\xi}_l, \label{LE-S} \\
\frac{d {\mathbf r}^{\rm L}_{l'}}{dt} & = & \left<\dot x^{\rm L}\right> \hat{\mathbf e}_x + \frac{1}{\zeta^{\rm L}}\sum_{l''} {\mathbf F}_{\rm dd}^{\rm LL}({\mathbf r}_{l'l''}) + \sqrt{2 D^{\rm L} } \,\grb{\xi}_{l'}, \label{LE-L} 
\end{eqnarray}
where $D^{\rm S}=k_{\rm B}T/\zeta^{\rm S}$ and $D^{\rm L}=k_{\rm B}T/\zeta^{\rm L}$
are diffusion constants. The dipolar force ${\mathbf F}_{\rm dd}^{\rm LL}({\mathbf r}_{l'l''})=-\partial_{{\mathbf r}_{l'}}U_{\rm dd}^{\rm eff}({\mathbf r}_{l'l''})$
is evaluated from the potential in Eq.~(\ref{Udd-eff}) with the only nonvanishing coupling strength
$\gamma^{\rm LL}= (\chi^{\rm L} V^{\rm L})^2 \mu_0/(8\pi)$.

\begin{figure}
\begin{center}
\resizebox{\columnwidth}{!}{\includegraphics{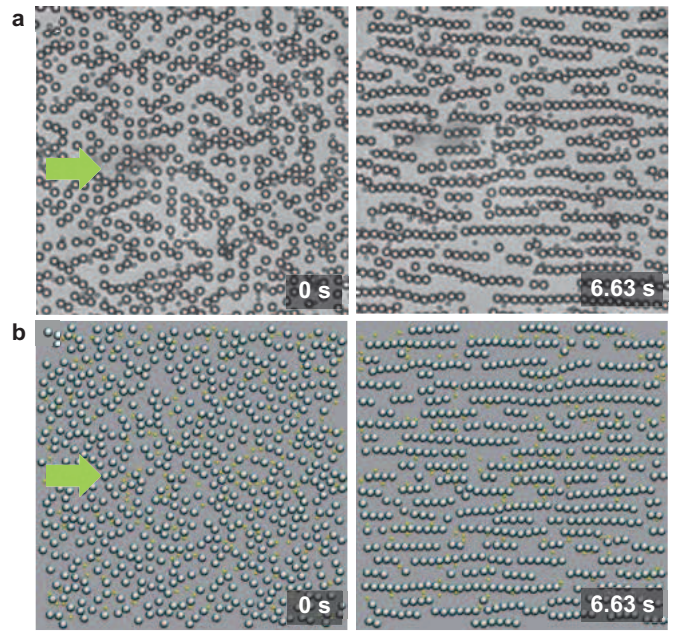}}
\caption{(a) Optical microscope images
separated by $6.63 \, {\rm s}$ showing the
formation of chains between only the
large particles starting from an initially random mixture.
The applied field has frequency $f=30\, {\rm Hz}$,
amplitude $H_0=780 \, {\rm A/m}$
and ellipticity $\beta=-0.12$,
\textit{i.e.} the same values used in the calculation of the potential in Fig.~\ref{fig:3}.
The particle net motion occurs from left to right as indicated by the green arrow.
(b) Corresponding snapshot obtained from
Brownian dynamics simulation of a binary mixture of
paramagnetic particles driven above the
FGF. The simulation based on Eqs.~(\ref{LE-S}) and (\ref{LE-L}) is performed for the same
values of parameters as the experiments shown in (a),
and the two snapshots are separated by $6.63\, {\rm s}$ in time. The corresponding video
can be found in the Supporting Information (VideoS1.wmv).}
\label{fig:4}
\end{center}
\end{figure}

Finally, we compare the experimental and simulation results
on the selective chaining process in Fig.~\ref{fig:4}.
In the top panel of this image,
we show two snapshots of a binary mixture of particles
driven above the FGF
by an elliptically polarized field
with field parameters, $f=30 \, {\rm Hz}$, $H_0=780 \,{\rm A/m}$
and $\beta=-0.12$. Starting from a random distribution of particles,
after $6.63 \, {\rm s}$ the large paramagnetic particles
form long chains aligned across the landscape which propel
to the right (along the longitudinal direction) at a certain speed, showing small
thermal fluctuation in the perpendicular direction.
In contrast, the small particles
are transported individually between the formed chains with the maximal speed $v_{\rm m}$,
without assembling in
any structure during their excursion.
The visual comparison with the numerical
simulation using
the reduced model, Eqs.~(\ref{LE-S}) and (\ref{LE-L}),
is displayed on the bottom panel of Fig.~\ref{fig:4}.
In both cases we observe the same type of dynamics
for both kinds of particles,
with a clear selective process in chain formation
arising from the difference in the
interparticle interactions.
Even by increasing the observation time,
we never detect any signature of chaining for the small particles,
which is attributed to trapping by the
travelling potential, dominating over the effects of the dipolar interactions.
The corresponding video, illustrating the
dynamics of the binary mixture of particles
in the experiment and simulation,
can be found in the Supporting Information (VideoS1.wmv).

\section{Conclusions}

To summarize, we have studied the dynamics of a binary mixture
of interacting paramagnetic colloidal particles in a travelling spatially periodic potential.
A sinusoidal landscape is generated by using
a magnetic structured substrate. Being additionally subjected
to a temporal modulation in the form of a rotating magnetic field,
the periodic landscape starts to move with the speed proportional to the
frequency of modulation, which causes the particles to propel.
Upon suitable choice of the field
parameters, we show that selective
chaining process can be induced by
forcing one type of particle
to assemble into condensed structures
which laterally confine
the motion of the smaller particles.

The selective chaining process
can be used to generate
remotely controllable
colloidal channels, confining the flow
of smaller magnetic particles~\cite{Ski13},
which presents a mesoscopic model system
for transport in a microfluidic medium.
Furthermore, of interest is the behaviour of the system at higher densities
and in particular how different relative fractions of particles affect transport characteristics.
More in general,
binary mixture of particles driven by
an external field have been investigated
theoretically~\cite{Dzu02,Dzu022,Rei06,Wys09,Spe09,Fri16}
and experimentally~\cite{Yel07,Vis11}
by different research groups.
Possible applications include
size segregation of magnetic species, fractionation in
lab-on-a-chip devices or transport of biological
species in analytic devices.

\begin{acknowledgement}
P. T. acknowledges support from the ERC Starting Grant "DynaMO" (no. 335040),
from Mineco (No. RYC-2011-07605
and No. FIS2013-41144-P) and AGAUR (Grant No.
2014SGR878).
A. V. S. and P. T. acknowledge support from a bilateral German-Spanish
program of DAAD (project no. 57049473) via the
Bundesministerium f\"ur Bildung und Forschung (BMBF).
A. V. S. acknowledges
P. T. and the Departament de F\'isica de la Mat\`eria Condensada for hosting
him as a visiting scientist at the University of Barcelona.
\end{acknowledgement}


\begin{thebibliography}{}

\bibitem{Mir05}
E. Mirowski, J. Moreland, A. Zhang, S. E. Russek, M. J. Donahue, Appl. Phys. Lett. \textbf{86}, 243901 (2005).

\bibitem{Ram06}
Q. Ramadana, Y. Chen, V. Samper, D. P. Poenar, Appl. Phys. Lett. \textbf{88}, 032501 (2006).

\bibitem{Gun05}
K. Gunnarsson, P. E. Roy, S. Felton, J. Pihl, P. Svedlindh, S. Berner, H. Lidbaum, S. Oscarsson, Adv. Mater. \textbf{17}, 1730 (2005).

\bibitem{Yel05}
B. B. Yellen, O. Hovorka and G. Friedman, Proc. Natl. Acad. Sci. U. S. A. \textbf{102}, 8860 (2005).

\bibitem{Tie07}
P. Tierno, S. V. Reddy, T. H. Johansen and T. M. Fischer, Phys. Rev. E \textbf{75}, 041404 (2007).

\bibitem{Don10}
M. Donolato, P. Vavassori, M. Gobbi, M. Deryabina, M. F. Hansen, V. Metlushko, B. Ilic, M. Cantoni, D. Petti, S. Brivio, R. Bertacco, Adv. Mater. \textbf{22}, 2706 (2010).

\bibitem{Hen11}
T. Henighan, D. Giglio, A. Chen, G. Vieira, and R. Sooryakumar, Appl. Phys. Lett. \textbf{98}, 103505 (2011).

\bibitem{Ehr11}
A. Ehresmann, D. Lengemann, T. Weis, A. Albrecht, J. Langfahl-Klabes, F. Gollner, and D. Engel,  Adv. Mater. \textbf{23}, 5568 (2011).

\bibitem{Gijs10}
M. A. Gijs, F. Lacharme and U. Lehmann, Chem Rev. \textbf{110}, 1518 (2010).

\bibitem{Mar13}
J. E. Martin, A. Snezhko, Rep. Prog. Phys. \textbf{76}, 126601 (2013).

\bibitem{Zbo99}
M. Zborowski, J. J. Chalmers, Magnetophoresis: Fundamentals and Applications,
Wiley Encyclopedia of Electrical and Electronics Engineering, 1�23 (2015).

\bibitem{Suw11}
M. Suwa, H. Watarai, Anal Chim Acta. \textbf{690}, 137 (2011).

\bibitem{Hel05}
L. E. Helseth, T. Backus, T. H. Johansen, and T. M. Fischer, Langmuir \textbf{21}, 7518 (2005).

\bibitem{Tie08}
P. Tierno, T. M. Fischer, T. H. Johansen, F. Sagu\'es, Phys. Rev. Lett. \textbf{100}, 148304 (2008).

\bibitem{Tie10}
P. Tierno, P. Reimann, T. H. Johansen, F Sagu\'es, Phys. Rev. Lett. \textbf{105}, 230602 (2010).

\bibitem{Tie20122}
P. Tierno, Phys. Rev. Lett., \textbf{109}, 198304 (2012).

\bibitem{Str2013}
A. V. Straube, P. Tierno, Europhys. Lett., \textbf{103}, 28001 (2013).

\bibitem{Tie081}
P. Tierno, S. V. Reddy, M. G. Roper, T. H. Johansen, T. M. Fischer, J. Phys. Chem. B \textbf{112}, 3833 (2008).

\bibitem{Tie082}
P. Tierno, A. Soba, T. H. Johansen, F. Sagu\'es, Appl. Phys. Lett. \textbf{93}, 214102 (2008).

\bibitem{Yel07}
B. B. Yellen, R. M. Erb, H. S. Son, R. Hewlin Jr., H. Shang, and G. U. Lee, Lab Chip \textbf{7}, 1681 (2007).

\bibitem{TiePCCP}
P. Tierno, F. Sagues, T. H. Johansen, T. M. Fischer, Phys. Chem. Chem. Phys. \textbf{11}, 9615 (2009).

\bibitem{Cli2007}
L. Clime, B. L. Drogoff, T. Veres, IEEE Trans. Magn. \textbf{43}, 2929 (2007).

\bibitem{Hel2007}
L. E. Helseth, J. Phys. D: Appl. Phys. \textbf{40}, 3030 (2007).

\bibitem{Tie2012}
P. Tierno, Soft Matter \textbf{8}, 11443 (2012).

\bibitem{Cro96}
J. C. Crocker, D. G. Grier, J. Colloid Interface Sci., \textbf{179}, 298 (1996).

\bibitem{Car15}
H.~Carstensen, V.~Kapaklis and M.~Wolff, Phys. Rev. E, \textbf{92}, 012303 (2015).

\bibitem{Lacis1997}
S. Lacis, J. C. Bacri, A. Cebers and R. Perzynski, Phys. Rev. E \textbf{55}, 2640 (1997).

\bibitem{Str2014}
A. V. Straube, P. Tierno, Soft Matter \textbf{10}, 3915 (2014).

\bibitem{Fer16}
F. Martinez-Pedrero, P. Tierno, T. H. Johansen, A. V. Straube, Scientific Reports \textbf{6}, 19932 (2016).

\bibitem{Ski13}
T. O. E. Skinner, S. K. Schnyder, D. G. A. L. Aarts, J. Horbach, and R. P. A. Dullens, Phys. Rev. Lett. \textbf{111}, 128301 (2013).

\bibitem{Dzu02}
J. Dzubiella, G. P. Hoffmann, H. L\"owen, Phys. Rev. E  \textbf{65}, 021402 (2002).

\bibitem{Dzu022}
J. Dzubiella, H. L\"owen, J. Phys.: Condens. Matter  \textbf{14}, 9383 (2002).

\bibitem{Rei06}
C. Reichhardt, C. J. Olson Reichhardt, Phys. Rev. E  \textbf{74}, 011403 (2006).

\bibitem{Wys09}
A. Wysocki, H. L\"owen, Phys. Rev. E  \textbf{79}, 041408 (2009).

\bibitem{Spe09}
D. Speer, R. Eichhorn, P. Reimann, Phys. Rev. Lett.   \textbf{105}, 090602 (2009).

\bibitem{Fri16}
R. Lugo-Frias, S. H. L. Klapp, J. Phys. Condensed Matter (2016), in press (arXiv:1511.07691).

\bibitem{Vis11}
T. Vissers, A. van Blaaderen, A. Imhof, Phys. Rev. Lett. \textbf{106}, 228303 (2011).

\end{thebibliography}
\end{document}